\documentclass{appolb}
\usepackage{graphicx}
\usepackage[fleqn]{amsmath}
\usepackage{amsfonts}
\usepackage{dcolumn}
\usepackage{bm}
\usepackage{braket}
\usepackage{multirow} 
\usepackage{MnSymbol}

\newcommand{\intt}
{
	\displaystyle \int
}

\newcommand{\diffn}[3]
{
	\frac{ { \partial }^{ #3 } #1 }{ \partial { #2 }^{ #3 } }
}

\begin{document}
\preprint{ }

\title{Astrophysical $S$-factor for $^6$Li$(p,\gamma)$$^7$Be in the coupled-channel Gamow shell model}
%
%

\author{G.X. Dong
\address{Grand Acc\'el\'erateur National d'Ions Lourds (GANIL), CEA/DSM - CNRS/IN2P3,
BP 55027, F-14076 Caen Cedex, France}  
\vskip 0.6truecm  
K. Fossez
\address{NSCL/FRIB Laboratory, Michigan State University, East Lansing, Michigan 48824, USA}  
\vskip 0.6truecm    
N. Michel, M. P{\l}oszajczak
\address{Grand Acc\'el\'erateur National d'Ions Lourds (GANIL), CEA/DSM - CNRS/IN2P3,
BP 55027, F-14076 Caen Cedex, France}
}
\maketitle

\begin{abstract}
{
  We have applied the Gamow shell model (GSM) in the coupled-channel representation to study the astrophysical $S$-factor for the proton radiative capture reaction of $^6$Li. Reaction channels are built by coupling the proton wave function expanded in different partial waves with the GSM wave functions of the ground state ($1^+$) and the excited states ($3^+_1$, $0^+_1$ and $2^+_1$) of $^6$Li. All relevant $E1$, $M1$, and $E2$ transitions from the initial continuum states in $^7$Be to the final bound states (${3/2}^-_1$ and ${1/2}^-_1$) states are included. It is found that the calculated total astrophysical $S$ factor for this reaction agrees well with the experimental data. }
\end{abstract}

\section{Introduction}
\label{intro}
The structure of low-lying states in nuclei from the valley of beta-stability valley is well
described by the standard shell model (SM). In a vicinity of the neutron or proton drip lines, atomic nuclei become weakly bound or even unbound in the ground state and, hence the description of its basic properties requires an explicit including of the coupling to the scattering continuum. The comprehensive description of bound states, resonances and scattering many-body states is possible only in either the continuum shell model~\cite{rf:7,rf:9} or the Gamow shell model (GSM)~\cite{Michel02,Michel03,Michel09}.

GSM is the rigged Hilbert space generalisation of the SM~\cite{Michel09}. Many-body states are expanded in the basis of Slater determinants spanned by bound, resonance and (complex-energy) non-resonant scattering states of the complete single particle (s.p.) Berggren ensemble~\cite{rf:4}. In the past, GSM has been applied successfully to describe structural properties of many-body bound states, resonances and their decays. In order to describe both the nuclear structure and reactions in a unified theoretical framework, the GSM has been recently formulated in the coupled-channel (CC) representation~\cite{Jaganathen14,Fossez15}. The GSM-CC approach has been applied to the low-energy elastic and inelastic proton scattering~\cite{Jaganathen14} and proton/neutron radiative capture reactions~\cite{Fossez15}.

The low-energy proton radiative capture reactions play an important role in the nuclear astrophysics, in particular in the nucleosynthesis of light end medium-heavy elements. In recent years, much interest has been devoted to the study of reactions which can produce $^7$Be in the stellar environment~\cite{Adelberger11,Angulo99}, especially to the $^6$Li$(p,\gamma)$$^7$Be reaction which is crucial for the consumption of $^6$Li and the formation of $^7$Be. This reaction can contribute to the understanding of the solar neutrino problem and pp-II, pp-III reaction chains since it produces $^7$Be which is destroyed by the $^7$Be$(p,\gamma)$$^8$B reaction. In this work, we apply the microscopic GSM-CC approach with the translationally invariant Hamiltonian and the effective finite-range two-body interaction to study the low-energy astrophysical $S$-factors in the reaction $^6$Li$(p,\gamma)$$^7$Be.

\section{The Gamow shell model in the coupled-channel representation}
\label{sec-1}

\subsection{The Hamiltonian}
\label{sec-2}

In our studies, the GSM Hamiltonian is written in the intrinsic nucleon-core coordinates of the cluster-orbital shell model~\cite{Michel09}:
\begin{equation}
	\hat{H} = \sum_{i = 1}^{ {N}_{ \text{val} } } \left( \frac{ \hat{\vec{p}}_{i}^{2} }{ 2 { \mu }_{i} } + {U}_{c} ( \hat{r}_{i} ) \right) + \sum_{i < j}^{ {N}_{ \text{val} } } \left( V ( \hat{\vec{r}}_{i} - \hat{\vec{r}}_{j} ) + \frac{ {\hat{\vec{p}}_{i}}{\cdot} {\hat{\vec{p}}_{j} }}{ {M}_{c} } \right),
	\label{GSM_Hamiltonian}
\end{equation}
where $N_\text{val}$ is the number of valence nucleons, $M_c$ is the mass of the core, $\mu_i$ is the reduced proton/neutron mass, $U_c(\hat{r})$ is the single-particle potential which describes the field of a core acting on each nucleon. The last term in Eq.~(\ref{GSM_Hamiltonian}) represents the recoil term, and $V(\hat{\vec{r}}_i-\hat{\vec{r}}_j)$ is the two-body interaction between valence nucleons.

\subsection{The GSM coupled-channel equations}
\label{sec-3}

The antisymmetric eigenstates of GSM-CC equations can be expanded in the complete basis of channel states ( ${ \ket{ r , c } = \hat{ \mathcal{A} } ( \ket{r} \otimes \ket{c}  }$):
\begin{equation}
\hat{ \mathcal{A} }\ket{ \Psi } = \ket{ \Psi } = \sumint\limits_{c} \intt_{0}^{ \infty } dr \, {r}^{2} \frac{ {u}_{c} (r) }{r} \ket{ r , c }  \ ,
	\label{eq_expan_eig_full_channel_basis}
\end{equation}
where ${ \hat{ \mathcal{A} } }$ is the antisymmetrization operator, ${u}_{c} (r) /r$ are the antisymmetrized channel wave functions: ${u}_{c} (r) /r \equiv \braket{ r , c | \Psi } \label{eq_as_channel_wfs}$.
Inserting Eq.~\eqref{eq_expan_eig_full_channel_basis} in the Schr\"odinger equation and then projecting it onto a given channel basis state ${ \bra{ r' , c' } }$, one obtains the GSM-CC equations:
\begin{equation}
	\sumint\limits_{c} \intt_{0}^{ \infty } dr \, {r}^{2} \left( {H}_{ c' , c } ( r' , r ) - E {N}_{ c' , c } ( r' , r ) \right) \frac{ {u}_{c} (r) }{r} = 0 \ ,
	\label{eq_CC_eqs_general}
\end{equation}
where ${H}_{ c' , c } ( r' , r ) = \braket{ r' , c' | \hat{H} | r , c }$ and
${N}_{ c' , c } ( r' , r ) = \braket{ r' , c' | r , c }$
are the Hamiltonian matrix elements and the norm matrix elements in the channel representation, respectively.

The channel state $\ket{ r , c }$ can be constructed using a complete Berggren set of s.p. states~\cite{rf:4}:
\begin{equation}
	\ket{ r , c } = \sum_{i} \frac{ {u}_{i} (r) }{r} \ket{ { \phi }_{i}^{ \text{rad} } , c } \ ,
	\label{eq_CC_basis_state_expansion_Berggren_2}
\end{equation}
where ${ \ket{ { \phi }_{i}^{ \text{rad} } , c } = \hat{ \mathcal{A} } ( \ket{ { \phi }_{i}^{ \text{rad} } } \otimes \ket{c} ) }$, ${u}_{i} (r) / r  = { \braket{ { \phi }_{i}^{ \text{rad} } | r }}$, and $\ket{ { \phi }_{i}^{ \text{rad} }}$ is the radial part.
The antisymmetry between the low-energy target states and the high-energy projectile states can be neglected in most cases. Hence, the expansion
~\eqref{eq_CC_basis_state_expansion_Berggren_2} can be written as
\begin{align}
	\ket{ r , c } &= \sum_{ i = 1 }^{ {i}_{ \text{max} } - 1 } \frac{ {u}_{i} (r) }{r} \ket{ { \phi }_{i}^{ \text{rad} } , c }
	+ \ket{r} \otimes \ket{c} - \sum_{ i = 1 }^{ {i}_{ \text{max} } - 1 } \frac{ {u}_{i} (r) }{r} \ket{ { \phi }_{i;c_{\rm proj}} } \otimes \ket{ {c}_{ \text{targ} } } \ ,
	\label{xxx}
\end{align}
where ${ {i}_{ \text{max} } }$ denotes the index from which the antisymmetry effects are neglected, and  $\ket{ { \phi }_{i;c_{\rm proj}} }$, $\ket{ {c}_{ \text{targ} }}$ are the projectile and target states, respectively.
Then, using the expansion~\eqref{xxx}, one can calculate the matrix elements of the Hamiltonian:
\begin{eqnarray}
	{H}_{ c' , c } ( r' , r ) = &-& \frac{ { \hbar }^{2} }{ 2 \mu } \left( \frac{1}{r} \diffn{ ( r \cdot ) }{r}{2} - \frac{ l ( l + 1 ) }{ {r}^{2} } - {k}_{ {c}_{ \text{targ} } }^{2} \right)
	\times \frac{ \delta ( r - r' ) }{ {r}^{2} } { \delta }_{ {c}_{ \text{targ} }' , {c}_{ \text{targ} } } \nonumber \\
	&+& {V}_{ c' , c } ( r' , r ) \ ,
	\label{eq_CC_EM_H_final}
\end{eqnarray}
where
${k}_{ {c}_{ \text{targ} } }^{2} = 2 \mu {E}_{ {c}_{ \text{targ} } } /{ { \hbar }^{2} }$ and
the channel-channel coupling potential ${V}_{ c' , c } ( r' , r )$ is given by:
\begin{equation}
	{V}_{ c' , c } ( r' , r ) = {U}_{ \text{basis} } (r) \frac{ \delta ( r - r' ) }{ {r}^{2} } { \delta }_{ {c}_{ \text{targ} }' , {c}_{ \text{targ} } } + \tilde{V}_{ c' , c } ( r' , r )
	\label{eq_CC_ME_Vccp_pot}
\end{equation}
with
\begin{eqnarray}
	\tilde{V}_{ c' , c } ( r' , r ) &=& \sum_{i,i' = 1 }^{{i}_{ \text{max} }} \frac{ {u}_{ i' } ( r' ) }{ r' } \frac{ {u}_{i} (r) }{r} {H}_{ c' , c } ( i' , i ) \nonumber \\
	 &-& \sum_{ i = 1 }^{ {i}_{ \text{max} } - 1 } \frac{ {u}_{ i } ( r' ) }{ r' } \frac{ {u}_{i} (r) }{r} ( {E}_{ i , {c}_{ \text{proj} } } + {E}_{ {c}_{ \text{targ} } } ){ \delta }_{ {c}_{ \text{targ} }' , {c}_{ \text{targ} } }  \ .
	\label{eq_CC_EM_Vccp_pot_rest}
\end{eqnarray}

In general, different channel states $\ket{ r , c }$ are nonorthogonal what leads to the generalized eigenvalue problem. To solve it, we use the orthogonal channel basis $\ket{ r , c }_{o} = \hat{O}^{ -\frac{1}{2} } \ket{ r , c } \label{eq_CC_non_ortho_to_ortho_channel_states}$:
\begin{equation}
	{}_{o}\braket{ r' , c' | r , c }_{o} = \frac{ \delta ( r' - r ) }{ {r}^{2} } { \delta }_{ c' c }
	\label{eq_CC_ortho_channel_basis_braket}
\end{equation}
The GSM-CC equations~\eqref{eq_CC_eqs_general} in this basis become:
\begin{align}
	\sumint\limits_{c} \intt_{0}^{ \infty } dr \, {r}^{2} &\left( {{}_{o}}\braket{ r' , c' | \hat{H}_{o} | r , c }_{o} - E {}_{o}\braket{ r' , c' | \hat{O} | r , c }_{o} \right){}_{o}\braket{ r , c | { \Psi }_{o} } = 0 \ ,
	\label{eq_CC_eqs_general_clear_ortho}
\end{align}
where
${ \hat{H}_{o} = \hat{O}^{  \frac{1}{2} } \hat{H} \hat{O}^{  \frac{1}{2} } }$, and $ {\ket{ \Psi_o } = \hat{O}^{1/2} \ket{ \Psi } }$. With a substitution: ${ \ket{ \Phi } = \hat{O} \ket{ \Psi } }$, this generalized eigenvalue problem can be transformed into a standard one:
\begin{equation}
	\sumint\limits_{c} \intt_{0}^{ \infty } dr \, {r}^{2} ( {}_{o}\braket{ r' , c' | \hat{H} | r , c }_{o} - E {}_{o}\braket{ r' , c' | r , c }_{o} ) {}_{o}\braket{ r , c | \Phi } = 0.
	\label{eq_CC_eqs_general_clear_ortho_again}
\end{equation}

Back in the nonorthogonal channel basis, these CC equations become:
\begin{equation}
	\sumint\limits_{c} \intt_{0}^{ \infty } dr \, {r}^{2} \braket{ r' , c' | \hat{H}_{m} | r , c } \frac{ {w}_{c} (r) }{r} = E \frac{ {w}_{ c' } ( r' ) }{ r' } \ ,
	\label{eq_CC_final}
\end{equation}
where ${w}_{c} (r)/ r \equiv \braket{ r , c | \hat{O}^{ \frac{1}{2} } | \Psi } = {}_{o}\braket{ r , c | \Phi }$, and ${ \hat{H}_{m} = \hat{O}^{ - \frac{1}{2} } \hat{H} \hat{O}^{ - \frac{1}{2} } }$ is the modified Hamiltonian.
Detailed discussion of the method of solving the integro-differential GSM-CC equations can be found in Ref. \cite{Fossez15}. Therein, one may also find all details of the proton/neutron radiative capture cross-section calculation in the GSM-CC approach.

\section{Discussion of results}

In the present calculations, we choose $^4$He as the core. For each considered partial wave: $l$=0, 1, and 2, the potential generated by the core is described by the Woods-Saxon (WS) potential with a spin-orbit term and the Coulomb potential of radius $r_c$=2.5 fm. For a two-body force, we use the Furutani-Horiuchi-Tamagaki (FHT) finite-range two-body interaction~\cite{Furutani79}. GSM and GSM-CC calculations are performed in two resonant shells: $0p_{3/2}$ and $0p_{1/2}$, and several shells in the non-resonant continuum on discretized contours: $\mathcal{L}^+_{s_{1/2}}$, $\mathcal{L}^+_{p_{1/2}}$, $\mathcal{L}^+_{p_{3/2}}$, $\mathcal{L}^+_{d_{3/2}}$, and $\mathcal{L}^+_{d_{5/2}}$. To reduce the size of the GSM Hamiltonian matrix, the basis of Slater determinants is truncated by limiting the occupation of $p_{3/2}$, $p_{1/2}$, $s_{1/2}$, $d_{3/2}$ and $d_{5/2}$ scattering states to two particles.

By adjusting parameters of the WS core potential and the FHT two-body interaction (for more details see Ref.~\cite{Fossez15}), the binding energies of low-lying states and the proton separation energies of $^6$Li and $^7$Be are well reproduced in the GSM.
The reaction channels in GSM-CC calculations are obtained by coupling the ground state $1^+$ and the excited states $3^+_1$, $0^+_1$ and $2^+_1$ of $^6$Li with the proton partial waves: $s_{1/2}$, $p_{1/2}$, $p_{3/2}$, $d_{3/2}$ and $d_{5/2}$. The channel-channel coupling potentials in GSM-CC have been slightly adjusted for ${3/2}_1^-$ and ${1/2}_1^-$ states of $^7$Be to compensate for the missing correlations due to neglected non-resonant channels. In the calculation of electromagnetic transitions, we take the standard effective charges for proton and neutron in the case of $E1$ and $E2$ transitions~\cite{Hornyak75}, while there are no effective charges for $M1$ transitions.

\begin{figure}[htb]
\centering
\includegraphics[width=13cm,clip]{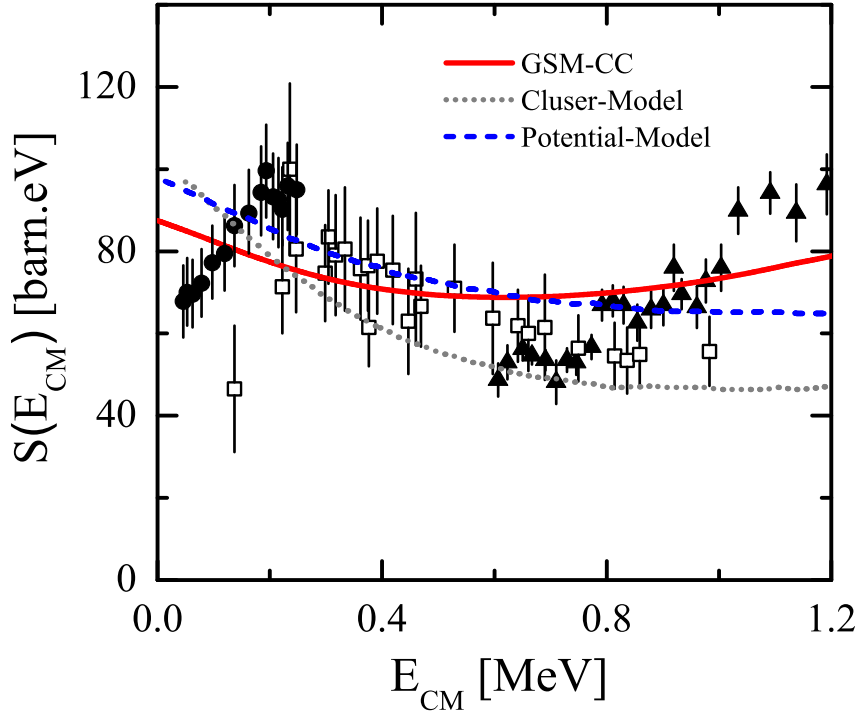}
	\caption{(Color online) Plot of the total astrophysical factor for the $^6\text{Li}(p,\gamma)^7\text{Be}$ reaction. Data are taken from ~\cite{Switkowski79} (open squares), \cite{He13} (filled circles), and~\cite{Xu13} (filled triangles). The solid line represents the exact, fully antisymmetrized GSM-CC calculation for capture to both the ground state $J^{\pi}=3/2_1^-$ and the first excited state $J^{\pi}=1/2_1^-$ of ${ {}^{7}\text{Be}}$. Calculations by the potential model~\cite{Huang10} and the microscopic cluster model~\cite{Arai02} are shown with the dashed and dotted lines, respectively.}
	\label{fig-1}
\end{figure}

All relevant $E1$, $M1$, $E2$ transitions from the initial continuum states $J_i = {3/2}^-, {1/2}^-, {5/2}^-$ in $^7$Be to the final bound state $J_f = {3/2}_1^-, {1/2}_1^-$ states have been included in the calculation of astrophysical $S$ factor. GSM-CC results are shown with the solid line in Fig. \ref{fig-1}. These results are compared both with the experimental data~\cite{Switkowski79,He13,Xu13} and other theoretical studies~\cite{Arai02,Huang10}.
Switkowski et al.~\cite{Switkowski79} measured $^6$Li$(p,\gamma)$$^7$Be cross section over a wide energy range of astrophysical interest. These data are well described by the GSM-CC calculations except for the lowest experimental point at $E_\text{CM}=$ 140 keV. GSM-CC results agree also qualitatively with other theoretical studies~\cite{Barker80,Arai02,Huang10} but predict a lower value for $S(0)$. One may notice that above $E_{\rm CM} = 1$ MeV, the agreement with the data deteriorates. This may due to the absence of higher-lying discrete and continuum states of $^6$Li target in the channel basis.

Recently, He et al. \cite{He13} reported a sudden drop of the astrophysical  factor $S(E_{\rm CM})$ at low energies and predicted a new positive parity resonance, ${1/2}^+$ or ${3/2}^+$, at $E_\text{CM}\simeq 0.195$ MeV. We do not confirm this finding in the GSM-CC calculations (see Fig.~\ref{fig-1}).

The energy dependence of the astrophysical $S$ factors has been studied by Prior et al.~\cite{Prior04} who showed that $S(E_{\rm CM})$ has a negative slope towards low energies, while an earlier measurement indicated a positive slope~\cite{Cecil92}. In our studies, the slope of $S(E_{\rm CM})$ is negative, and $S_{\rm GSM-CC}(0)$=88.3423 b$\cdot$eV is close to the accepted experimental value $S_{\rm exp}(0)=$79$\pm$18 b$\cdot$eV.

In conclusion, the GSM-CC calculations provide a comprehensive and unified description of both the low-energy spectra of $^6$Li and $^7$Be as well as the proton radiative capture cross section in the reaction $^6\text{Li}(p,\gamma)^7\text{Be}$. The same Hamiltonian is used both in the structure and in the reaction studies. The restriction of a number of excitations into states of the non-resonant continuum and the absence of reaction channels build from the non-resonant states in the continuum of $^6$Li both limit the configuration space in GSM-CC as compared to the GSM. This approximation in the GSM-CC wave function is corrected by a small adjustment of the channel-channel coupling potentials. The full account of our studies for mirror reactions $^6\text{Li}(p, \gamma)^7\text{Be}$, $^6\text{Li}(n,\gamma)^7\text{Li}$ and mirror nuclei $^7$Be, $^7$Li will be reported elsewhere \cite{future}.

\vskip 0.5truecm
One of us (G.X. Dong) wishes to thank for the support from the Helmholtz Association (HGF) through the Nuclear Astrophysics Virtual Institute (VH-VI-417).

\end{document}